\documentstyle[12pt,twoside,fleqn,espcrc1]{article}
\title{Thermal rho's in the quark-gluon plasma}
\author{Robert D. Pisarski\address{Department of Physics,
	Brookhaven National Laboratory,\\
	P.O. Box 5000, Upton, NY, 11973, USA \\}}
\begin{document}
\maketitle
\begin{abstract}
I discuss different models which predict changes in the
mass of the thermal $\rho$ field.  I emphasize that while
the predictions are strongly model dependent, nevertheless
substantial shifts in the thermal $\rho$ mass are expected
to occur at the point of phase transition.  As long as the
thermal $\rho$ peak does not become too broad, this should
provide a striking signature of the existence of a
phase transition.
\end{abstract}
\begin{center}
{\it Talk presented at Quark Matter '95}
\end{center}
\section{INTRODUCTION}
In this note I examine various models for how the
thermal $\rho$ peak might shift.  While the predictions
of these models are diametrically opposed, my intention
herein, and in general, is {\it not} to insist the predictions
of one model are necessarily superior to another.  If
the results of numerical simulations of lattice gauge
theory have taught us anything, they have demonstrated that
the phase diagram of $QCD$ is rather complicated and
intricate.  Instead of
insisting upon the virtues of a single model, it is
perhaps more reasonable to explore different models,
and the range of possible predictions.  I try to
stress what is most general in each case.

\section{DECONFINEMENT DOMINATES}

The simplest model of confinement is the bag model \cite{bag}.
If $b$ is the bag constant, then the energy of a hadron is
computed by carving out a cavity of radius $R$ from the
vacuum.  The energy is a balance between the volume energy
of the cavity, versus the zero point energy which arises
from confining quarks to the cavity:
$
E_i(R) = (4 \pi/3)b R^3 + a_i/R ,
$
where $a_i$ is a constant which depends upon the individual
hadron of type $i$.  Minimizing $E_i(R)$ with respect to $R$,
$E_i \sim b^{1/4}$, and all hadron masses are
proportional to $b^{1/4}$.  This is really a trivial consequence of
the model, since the only dimensional parameter is the bag constant,
$b$.

To characterize the effects of nonzero temperature it is
convenient to introduce
a temperature dependent bag constant, $b(T)$ \cite{rdp1}.
At zero
temperature the bag constant is
minus the pressure of the vacuum, pushing in upon the quarks
and gluons confined to the bag.  At nonzero temperature, then,
we include the pressure of quarks and gluons pushing out on the
bag, and pions pushing in on the bag, to obtain
$
b(T) = b - \pi^2 T^4 ( 8/45 + 7/60 - 1/30) .
$
Hence there is a deconfining phase transition when the
effective bag constant vanishes,
$
b(T_d) = 0.
$
Approaching the temperature of deconfinement from below,
the effective bag constant
vanishes linearly with temperature, $b(T) \sim T_d -T$.
Plugging this back into the equation for the energy of
a hadron and solving, the mass of a hadron --- any hadron --- vanishes like
$
m_i(T) \sim
(T_d - T)^{1/4} \; .
$

The conclusion that {\it all} hadron masses vanish at
$T_d$ \cite{rdp1} is really an inexorable consequence of the fact
that there is only one dimensional parameter in the bag
model.  Once we characterize the phase transition by
one of vanishing bag constant, of necessity the masses must
vanish at $T_d$, since any mass can only be proportional
to this single mass scale.

This was proposed as a general
principle by Brown and Rho \cite{brown}.  Even if the phase
transition is of first order, it is reminiscent of a second order transition.
The striking difference is that according to Brown-Rho scaling,
an infinite number of
hadrons becomes massless at $T_d$, while in a second order transition,
the number of massless fields is finite, equal to the
rank of the representation for the
the symmetry group appropriate for the transition.

Another model which gives similar behavior is
that of sum rules \cite{sum}.
At zero temperature the vacuum
is characterized by several condensates, including a
gluon condensate,
$\langle (G_{\mu \nu})^2 \rangle$, and a quark condensate,
$ \langle \bar{\psi} \psi \rangle$.
At nonzero temperature, in the chiral limit the quark condensate
$ \langle \bar{\psi} \psi \rangle$ vanishes above the chiral transition,
but the gluon condensate does not.  Also, at nonzero temperature there
are two gluonic condensates, for electric and magnetic fields,
$\langle (E_{i})^2 \rangle$ and
$\langle (B_{i})^2 \rangle$.
Since even at the point of phase transition there are dimensional
mass parameters about,
typically hadronic masses are nonzero at $T_d$.
Hadron masses do tend
to fall, usually dramatically, because the gluonic condensates
decrease with temperature (see, however, \cite{sumup}).

I characterize these models in which hadron masses scale more or less
uniformly with temperature as ones in which deconfinement dominates.
This description is admittedly imprecise.
In a theory without quarks, we
know how to specify the phase transition, by the breaking
of a global $Z(3)$ symmetry above $T_d$.  For such a phase
transition, the $Z(3)$ symmetry does not constrain how hadronic
(here glueball) masses scale as $T_d$ is approached.

In these models the $\rho$ mass goes down with increasing
temperature \cite{rdp1}, along with every other hadron mass scale.

\section{CHIRAL SYMMETRY RESTORATION DOMINATES}

A different approach is to forget about how confinement arises
in the first place, and concentrate upon the restoration of the
global chiral symmetry at a temperature $T_\chi$ \cite{rdp2}.
By its very nature, this approach will be limited to temperatures
at or below $T_\chi$.  For example, such models cannot, in any
elementary manner, incorporate the large increase in entropy
which numerical simulations of lattice gauge theory find is
a universal feature of the phase transition, irrespective of
values of the quark masses \cite{lattice}.

Unlike the previous section, with one dimensional parameter
and an almost universal parametrization of the hadron spectrum,
the characterization of chiral symmetry restoration requires
a precise specification of all fields in the proper chiral
multiplets.  Assuming that effects of the axial anomaly are always
significant, so that we need only classify particles according
to $SU(2) \times SU(2)$ multiplets, I introduce
$
\Phi = \sigma \, t^0 + i \vec{\pi} \! \cdot \! \vec{t} \; ,
$
where $\sigma$ is an isosinglet $0^+$ field.
For the left and right handed vector fields I take
$
A^\mu _{l,r} = (\omega^\mu \pm f_1^\mu)t^0
+ (\vec{\rho}^{\, \mu} \pm \vec{a}_1^{\, \mu})\!\cdot\! \vec{t}
$
where $\omega$ and $\vec{\rho}$ are $1^-$ fields,
and $f_1$ and $\vec{a}_1$ are $1^+$ fields.

The crucial assumption which I then make is that of vector meson
dominance.  This severely constrains the dimensionless couplings of
the model to be those which follow {\it exclusively} by promoting the
global chiral symmetry to a local chiral symmetry.  From the viewpoint of
effective lagrangians, this is really an utterly remarkable principle
\cite{rdp2}.  The effective Lagrangian is then a sum of terms
invariant under the local chiral symmetry,
\begin{equation}
{\cal L}_{VDM} = tr \left(\left|D^\mu \Phi\right|^2
- \mu^2 |\Phi|^2 + \lambda (|\Phi|^2)^2 - 2 h t^0 \Phi
+ (F_l^{\mu \nu})^2/2 + (F_r^{\mu \nu})^2/2 \right) \; ,
\end{equation}
and a single mass term which is solely responsible for the breaking
of the local chiral symmetry,
\begin{equation}
{\cal L}_{mass} =  m^2 \left( (A_l^\mu)^2 + (A_r^\mu)^2 \right)\; .
\end{equation}
Here $D_\mu$ and $F_{l,r}^{\mu \nu}$ are the appropriate covariant derivative
and field strengths for a local $SU(2) \times SU(2)$ symmetry: for instance,
$D^\mu \Phi = \partial^\mu \Phi - i g (A^\mu_l \Phi - \Phi A^\mu_r)$,
where $g$ is the coupling of vector meson dominance.

This effective lagrangian involves several parameters, but for the physics
of the $\rho$ mass, all we need to know is that at zero temperature,
$m^2_\rho = m^2$, and that
the $\rho-a_1$ mass degeneracy is lifted
after $\sigma$ acquires a vacuum expectation value
$=\sigma_0$,  $m^2_{a_1}
= m^2 + g^2 \sigma_0^2$.

Chiral symmetry predicts that since the $\rho$ and $a_1$ are chiral
partners under $SU(2)\times SU(2)$, that their masses should be equal
at the chiral transition (in the chiral limit).
This is all chiral symmetry says: it does not predict where the
$\rho$ and $a_1$ masses meet.

If one
looks at the expansion about zero temperature, the results are involved.
This is due to the fact that there is mixing between the $a_1$ and $\pi$
fields, which then mix the $\rho$ and $a_1$.  To order $T^2$,
according to a general analysis by Eletsky and Ioffe \cite{ei}, the thermal
$\rho$ and $a_1$ masses don't move at all; this is confirmed by studies
in the linear and nonlinear sigma models \cite{rdp2},\cite{ts}.
In the linear model, to order $T^4$, the $\rho$ mass goes down, and the
$a_1$, up \cite{rdp2}!  This is not universal: by a dispersive technique,
Eletsky and Ioffe find that both masses decrease, as in sum rules \cite{ei2}.

At the point of phase transition, however, I would argue that the nature
of the effective lagrangian, and {\it especially} the assumption of
vector meson dominance, alone
constrains the thermal $\rho$ mass to be greater than that at zero temperature
\cite{rdp2}.  The point is really trivial: the mass term above, $m^2$,
is by assumption independent of the scalar field, and so $\sigma_0$.  So
that part of the mass is fixed.  In addition, there are thermal fluctuations,
due to $\pi$'s, and, near the transition, $\sigma$'s, since those become
degenerate with the $\pi$'s at the chiral transition.  For scalar fields,
the effect of these fluctuations is inevitably to push the thermal $\rho$
mass {\it up} at the point of phase transition.  For example, an elementary
calculation to one loop order predicts that at $T_\chi$,
\begin{equation}
m_\rho(T_\chi)^2= m_{a_1}(T_\chi)^2 =
\frac{1}{3} \left( 2 m^2_\rho + m^2_{a_1} \right) = (962 \, MeV)^2 \; .
\end{equation}

If one abandons vector meson dominance, there is no unique prediction.
For example, if instead of the mass term above, suppose
that one insists that
the $\rho$ mass arises from spontaneous symmetry breaking.
This can be done by setting $m=0$ above, and adding the term
\begin{equation}
{\cal L}_\kappa = \kappa \; tr(|\Phi|^2) \;
tr ( (A_l^\mu)^2 + (A_r^\mu)^2 ) \; .
\end{equation}
where $\kappa$ is a dimensionless coupling constant.  At zero temperature,
$m_\rho^2 = \kappa \sigma_0^2$.  At nonzero temperature, the $\rho$ mass
decreases as the condensate evaporates.  Even so, it does not vanish
entirely, since thermal fluctuations from $\pi$'s and $\sigma$'s still
contribute.  A simple calculation \cite{rdp2} shows that for this term,
\begin{equation}
m^2_\rho(T_\chi) = m^2_{a_1}(T_\chi) = \frac{2}{3} m_\rho^2 =
(629 \, MeV)^2 \; .
\end{equation}
This is like the results from sum rules.

Consequently, and rather surprisingly, we see that the question of the
position of the thermal $\rho$ mass at the point of phase transition
provides a rather strong test of the applicability of the assumption
of vector meson dominance at nonzero temperature.  If vector meson
dominance holds, then the thermal $\rho$ mass is greater at $T_\chi$
than at zero temperature; if it does not hold, there is no unique
prediction, as the thermal $\rho$ mass could be either greater or less
than that at zero temperature.

\section{CONCLUSIONS}

Wherever the $\rho$ goes,
the really crucial question for experimentalists is how broad the
thermal $\rho$ peak becomes.  This is beyond the subject of present
analysis \cite{rdp2}; surely the thermal $\rho$ is broader than that
at zero temperature.  Pulling out the thermal $\rho$ peak will be
extremely difficult.  Nevertheless, if possible, it would provide
truly dramatic evidence for a new state of matter.

\end{document}